\documentclass[a4paper]{article}

\usepackage[english]{babel}
\usepackage[utf8x]{inputenc}
\usepackage[T1]{fontenc}

\usepackage[margin=1in]{geometry}

\usepackage{amsmath}

\usepackage{graphicx}
\usepackage[colorinlistoftodos]{todonotes}
\usepackage[colorlinks=true, allcolors=blue]{hyperref}
\mathchardef\mhyphen="2D
\usepackage{subfig}
\usepackage{latexsym,theorem}
\usepackage{amsbsy}
\usepackage{amssymb}
\usepackage{bm}
\usepackage{mathptmx}
\usepackage[normalem]{ulem}
\usepackage{xcolor}
\usepackage{booktabs}
\usepackage{float}
\usepackage{placeins}
\usepackage[title]{appendix}
\usepackage{natbib}
\usepackage{authblk}

\usepackage{scalerel,stackengine}
\stackMath
\newcommand\reallywidehat[1]{%
\savestack{\tmpbox}{\stretchto{%
  \scaleto{%
    \scalerel*[\widthof{\ensuremath{#1}}]{\kern.1pt\mathchar"0362\kern.1pt}%
    {\rule{0ex}{\textheight}}
  }{\textheight}%
}{2.4ex}}%
\stackon[-6.9pt]{#1}{\tmpbox}%
}

\newcommand\independent{\protect\mathpalette{\protect\independenT}{\perp}}
\def\independenT#1#2{\mathrel{\rlap{$#1#2$}\mkern2mu{#1#2}}}

\usepackage{color}
\usepackage{fancyvrb}

\DefineVerbatimEnvironment{Highlighting}{Verbatim}{commandchars=\\\{\}}
\usepackage{framed}
\definecolor{shadecolor}{RGB}{248,248,248}
\newenvironment{Shaded}{\begin{snugshade}}{\end{snugshade}}
\newcommand{\KeywordTok}[1]{\textcolor[rgb]{0.13,0.29,0.53}{\textbf{#1}}}
\newcommand{\DataTypeTok}[1]{\textcolor[rgb]{0.13,0.29,0.53}{#1}}
\newcommand{\DecValTok}[1]{\textcolor[rgb]{0.00,0.00,0.81}{#1}}

\newcommand{\FloatTok}[1]{\textcolor[rgb]{0.00,0.00,0.81}{#1}}

\newcommand{\StringTok}[1]{\textcolor[rgb]{0.31,0.60,0.02}{#1}}

\newcommand{\CommentTok}[1]{\textcolor[rgb]{0.56,0.35,0.01}{\textit{#1}}}

\newcommand{\OtherTok}[1]{\textcolor[rgb]{0.56,0.35,0.01}{#1}}

\newcommand{\ControlFlowTok}[1]{\textcolor[rgb]{0.13,0.29,0.53}{\textbf{#1}}}
\newcommand{\OperatorTok}[1]{\textcolor[rgb]{0.81,0.36,0.00}{\textbf{#1}}}

\newcommand{\NormalTok}[1]{#1}

\title{On variance of the treatment effect in the treated using inverse probability weighting}
\author[1,*]{Sarah A. Reifeis}
\author[1,**]{Michael G. Hudgens}
\affil[1]{Department of Biostatistics, University of North Carolina at Chapel Hill, Chapel Hill, NC, USA}
\affil[*]{sreifeis@email.unc.edu}
\affil[**]{mhudgens@email.unc.edu}
\date{}   
\begin{document}
\maketitle

\begin{abstract}

In the analysis of observational studies, inverse probability weighting (IPW) is commonly used to consistently estimate the average treatment effect (ATE) or the average treatment effect in the treated (ATT). The variance of the IPW ATE estimator is often estimated by assuming the weights are known and then using the so-called ``robust'' (Huber-White) sandwich estimator, which results in conservative standard error (SE) estimation. Here it is shown that using such an approach when estimating the variance of the IPW ATT estimator does not necessarily result in conservative SE estimates. That is, assuming the weights are known, the robust sandwich estimator may be conservative or anti-conservative. Thus confidence intervals of the ATT using the robust SE estimate will not be valid in general. Instead, stacked estimating equations which account for the weight estimation can be used to compute a consistent, closed-form variance estimator for the IPW ATT estimator. The two variance estimators are compared via simulation studies and in a data analysis of the effect of smoking on gene expression.

\end{abstract}

Keywords: confounding, estimating equations, exposure effect, Huber-White sandwich variance estimator, observational data, variance estimation

\section{Introduction}

Observational studies are often used to draw inference about the effect of a treatment (or exposure) on an outcome of interest, especially in settings where randomized trials are not feasible. Common estimands for these types of analyses are the average treatment effect (ATE) and the average treatment effect in the treated (ATT). These estimands answer different causal questions, so it is critical to first establish the motivation and goals of inference when choosing a target estimand. Smoking, for example, is a commonly studied exposure where investigators may be interested in the ATT \citep{Moodie2018}. With smoking as the exposure, the ATT considers the effect of smoking only among those who smoke. On the other hand, the ATE contemplates the counterfactual scenario where everyone in the population smokes versus when no one smokes, which may not be of interest from a public health perspective. Another common example comes from pharmacoepidemiology, where the effect of a certain drug in users of that drug can often be the most relevant estimand for public health research \citep{Brookhart2013, Taylor2013, Nduka2016}. The ATT has utility across a range of disciplines where there is interest in treatment effects among only those individuals who ultimately receive the treatment. Other contexts in which the ATT is often the target of inference include health behavior and policy \citep{Rawat2010, Austin2011, Boulay2014, Were2017}, ecology and environmental management \citep{Gross2011, Tamini2011, Ramsey2019}, criminology \citep{Apel2010, Morris2016, Widdowson2016}, and economics and public policy \citep{Heckman2001, Addai2014, Marcus2014, Abdia2017, Jawid2019}. 

Inverse probability weighting (IPW) is often used for estimation of treatment effects from observational data, where confounding is expected in general. IPW estimators of the ATE and ATT can be computed by regressing the outcome on the exposure using weighted least squares, where the weights are functions of estimated propensity scores. The variance of the IPW ATE estimator is often estimated by assuming the weights are known and then using the so-called ``robust'' (Huber-White) sandwich estimator, which results in conservative standard error (SE) estimation~\citep{Lunceford2004, VanderWal2011, Hernan2020}. Likewise, the variance of the IPW ATT estimator is sometimes estimated by assuming the weights are known~\citep{Brookhart2013, Pirracchio2016, Ramsey2019}. However, unlike the IPW ATE estimator, there is no theoretical justification for assuming the weights are known when estimating the variance of the IPW ATT estimator~\citep{Bodory2018}. Indeed, herein we prove that such an approach can produce either conservative or anti-conservative SE estimates for the IPW ATT estimator. Consequently, confidence intervals using this approach will not be valid in general. Instead, stacked estimating equations which account for the weight estimation can be used to compute a consistent, closed-form variance estimator for the IPW ATT estimator.

The remaining sections of this paper are organized as follows. Section \ref{sec: Mthds} describes the IPW estimator for the ATT, and the corresponding variance estimators. Four simple examples are provided showing that the robust variance estimator with the weights assumed known can be conservative or anti-conservative. In Section \ref{sec: Sim Study} simulation studies are conducted for the four example scenarios outlined previously, and the finite sample properties of the variance estimators are explored. In Section \ref{sec: Data An} the Metabolic Syndrome in Men (METSIM) cohort data are analyzed with the methods outlined in Section \ref{sec: Mthds}. Section \ref{sec: Discussion} concludes with a discussion of the implications of these findings. 
Derivations and R code for replicating the results in the main text are provided in the Appendices.

\section{Methods}\label{sec: Mthds}

\subsection{IPW ATT Estimator\label{ssec: IPW Est}}
Consider an observational study where the goal is to draw inference about the effect of a binary exposure $A$ on an outcome $Y$. For $a=0,1$, let $Y^a$ denote the potential outcome had, possibly counter to fact, the exposure level been $a$. Let $Y$ denote the observed outcome, such that $Y=A Y^1 + (1-A)Y^0$. The ATT, the estimand of interest, is defined as $ATT = \mu_1 - \mu_0$ where $\mu_a = E(Y^a | A=1)$ for $a=0,1$ denotes the mean potential outcome under treatment $a$ among the treated individuals. 

With observational data, there is potential for confounding because individuals are not randomized to exposure $A$. IPW can be used to adjust for confounding of the relationship between the exposure and outcome. The inverse probability weights are functions of the propensity scores, which are often estimated by fitting the logistic regression model 
\begin{equation}
     \mbox{logit}\{P(A=1|L)\} = \alpha_0 + \alpha_1^TL  \label{eqn: wt logreg CT}
\end{equation}
where $L$ represents a (column) vector of $J$ measured pre-exposure variables and $\alpha_1$ is a parameter vector of length $J$. Assume we observe $n$ independent and identically distributed (i.i.d.) copies of $(L,A,Y)$ denoted by $(L_i,A_i,Y_i)$ for $i=1,...,n$. Let $\alpha = (\alpha_0, \alpha_1^T)$ and let $\hat{\alpha}$ be the maximum likelihood estimator (MLE) of $\alpha$ obtained by fitting model (\ref{eqn: wt logreg CT}). Let $h(L; \alpha) = P(A=1| L) / P(A=0|L) =  \exp(\alpha_0 + \alpha_1^TL)$ and $h(L; \hat{\alpha}) = \exp(\hat{\alpha}_0 + \hat{\alpha}_1^TL)$. Then the IPW ATT estimator equals
\begin{align}
    \widehat{ATT} = \frac{\sum_i \hat{W}_i A_i Y_{i}}{\sum_i \hat{W}_i A_i} - \frac{\sum_i \hat{W}_i (1-A_i) Y_{i}}{\sum_i \hat{W}_i (1-A_i)} \label{eqn: ATT Est}
\end{align} 
where $\hat{W}_i = W(A_i, L_i; \hat{\alpha}) = A_i + (1 - A_i) h(L_i; \hat{\alpha})$ is the estimated weight for subject $i$~\citep{Sato2003}, and $\sum_i = \sum_{i=1}^n$. The two ratios in (\ref{eqn: ATT Est}) are sometimes referred to as Hajek or modified Horwitz-Thompson estimators~\citep{Hernan2020}. The IPW ATT estimator (\ref{eqn: ATT Est}) is consistent for $ATT$ under the following assumptions: stable unit treatment value assumption~\citep{Rubin1980_SUTVA}; positivity, i.e., $P(A=a | L = l)>0$ for $a=0,1$ and all $l$ where $dF_L(l) > 0$ and $F_L$ is the CDF of $L$; conditional exchangeability, i.e., $Y^a \independent A | L$ for $a=0,1$; and correct specification of the model for $A|L$. Note that no outcome model for $Y$ given $A$ or $L$ is assumed. 

A convenient way to compute (\ref{eqn: ATT Est}) using standard software entails fitting a simple linear regression model of $Y$ on $A$ by weighted least squares. The variance of (\ref{eqn: ATT Est}) is sometimes then estimated by assuming the weights are known and computing the robust or Huber-White (HW) sandwich variance estimator, which is easily computed in standard software (e.g., \textit{sandwich} in R, or the \textit{REG} procedure with the \textit{WHITE} option in the \textit{MODEL} statement in SAS). While computationally convenient, this estimator will not generally result in valid inference, as shown below. 

\subsection{Variance Estimators of the ATT Estimator\label{ssec: Var Est}}
The asymptotic distribution of the IPW ATT estimator (\ref{eqn: ATT Est}) can be derived using standard estimating equation theory. In particular, let
\begin{align}
    \psi(Y_i, A_i, L_i, \alpha, \mu) &= 
    \begin{pmatrix}
        \psi_{\alpha} (A_i, L_i, \alpha) \\
        \psi_1 (Y_i, A_i, L_i, \alpha, \mu) \\
        \psi_0 (Y_i, A_i, L_i, \alpha, \mu) \\
    \end{pmatrix}  
    = \begin{pmatrix}
        \{A_i - e(L_i; \alpha)\} (1, L_{i}^T)^T \\
        W(A_i, L_i; \alpha) A_i (Y_i - \mu_1) \\
        W(A_i, L_i; \alpha) (1 - A_i) (Y_i - \mu_0) \\
    \end{pmatrix} \label{eqn: ATT sEE}
\end{align}
where $\mu=(\mu_1, \mu_0)$, $\psi_{\alpha}$ denotes the $J+1$ vector of score functions from the log likelihood corresponding to model (\ref{eqn: wt logreg CT}), and $e(L_i; \alpha) = P(A_i=1|L_i) = h(L_i; \alpha)/\{1 + h(L_i; \alpha)\}$ is the propensity score. The functions $\psi_{1}$ and $\psi_{0}$ correspond to the first and second ratios of the ATT estimator (\ref{eqn: ATT Est}), respectively.

Let $\xi = (\alpha_0, \alpha_1^T, \mu_1, \mu_0)^T$ and let $\hat \xi = (\hat \alpha_0, \hat \alpha_1^T, \hat \mu_1, \hat \mu_0)^T$, where $\hat \xi$ solves the estimating equations $\sum_i \psi(Y_i, A_i, L_i, \alpha, \mu) = 0$ and $\widehat{ATT} = \hat \mu_1 - \hat \mu_0$ is the ATT estimator in (\ref{eqn: ATT Est}). Then under suitable regularity conditions~\citep{Stefanski2002}
\begin{equation*}
    \sqrt{n} (\hat{\xi} - \xi) \rightarrow^d N(0, V(\xi))
\end{equation*}
where $V(\xi) = \mathbb{A}(\xi)^{-1} \mathbb{B}(\xi) \{\mathbb{A}(\xi)^{-1}\}^T$ with $\mathbb{B}(\xi) = E\{\psi(Y, A, L, \xi)\psi(Y, A, L, \xi)^T\}$, $\mathbb{A}(\xi) = E\{-\Dot{\psi}(Y, A, L, \xi)\}$, and $\Dot{\psi}(Y, A, L, \xi) = \partial \psi(Y, A, L, \xi) / \partial \xi^T$. It follows from the Delta method that (\ref{eqn: ATT Est}) is consistent and asymptotically normal, i.e., 
\begin{align}
    \sqrt n \left(\widehat{ATT} - ATT \right) \xrightarrow{d} N(0, \Sigma).\label{eqn: ATT asy norm}
\end{align}
where $\Sigma = \nabla g(\xi)^T V(\xi) \nabla g(\xi)$ is the asymptotic variance of the ATT estimator, $\nabla g(\xi)^T = (0, 0_J^T, 1, -1)$, and $0_J$ is the 0 vector of length $J$. Let $\hat{\Sigma}$ denote the consistent estimator for $\Sigma$ obtained by substituting $\hat{V}(\hat{\xi})$ for $V(\xi)$ where the expectations in $V$ are replaced by their empirical counterparts and $\hat \xi$ is substituted for $\xi$. Then in large samples the variance of $\widehat{ATT}$ can be approximated by $\hat \Sigma/n$, which below is referred to as the stacked estimating equations (SEE) variance estimator.

The derivation above considers the usual scenario in observational studies where the weights are estimated. Now suppose instead that the weights are known, i.e., the values of the true weights $W_i= A_i + (1-A_i) h(L_i; \alpha)$ are known and therefore the propensity score need not be estimated. Let $\widehat{ATT}^*$ denote the estimator (\ref{eqn: ATT Est}) with $\hat W_i$ replaced with $W_i$. Then, similar to above, it is straightforward to show that $\widehat{ATT}^*$ is consistent and asymptotically normal with asymptotic variance
\begin{align}
    \Sigma^{*} = p_1^{-2} \left[E\{A_i (Y_i^1 - \mu_1)^2\} + E\{(Y_i^0 - \mu_0)^2h(L_i; \alpha)e(L_i; \alpha)\} \right]. \label{eqn: sig star}
\end{align}
where $p_1 = P(A=1)$. Let $\hat \Sigma^*$ represent the estimator for $\Sigma^*$ obtained by substituting $p_1$ and the expectations in (\ref{eqn: sig star}) with their empirical counterparts and substituting $\hat \mu$ for $\mu$, where $\alpha$ is known. Then $\hat \Sigma^*/n$ denotes the robust (Huber-White) sandwich variance estimator discussed at the end of Section \ref{ssec: IPW Est}. 

In Appendix A it is shown that 
\begin{align}
    \Sigma &= \Sigma^{*} +  p_1^{-2}(c_{11} + c_{22} - 2c_{12}) \label{eqn: Sigma}
\end{align}
where the explicit forms of $c_{km},\ k,m \in \{1,2\}$ are given in the Appendix. In general, the sign of the second term on the right side of the equality of (\ref{eqn: Sigma}) can be either positive or negative. Therefore, $\Sigma^*$ can be either larger or smaller than $\Sigma$, as shown via four simple examples in the next section. This suggests that using $\hat \Sigma^*$ may result in conservative or anti-conservative inference.

\subsection{Asymptotic Calculations}\label{sec: Asy Calc}

In this section $\Sigma^*$ and $\Sigma$ are compared for four simple data generating processes. Table \ref{tab:SimVars CT} contains variable definitions and relationships for the variable $L$, the exposure $A$, and the potential outcomes $Y^a$, in each of four examples. In scenarios (i) and (ii) the variable $L$ is binary, and in scenarios (iii) and (iv) $L$ is continuous (Normal). In all four scenarios, the exposure $A$ is binary and $Y^a$ is Normally distributed with a standard deviation of $0.5$. The marginal exposure probability $p_1$ and the population ATT value are also given in Table \ref{tab:SimVars CT}; these scenarios were chosen because they do not involve rare exposures or extreme effect sizes. 
 

\begin{table}[H]
    \centering
    \caption{Distribution of $L$, exposure $A$, and potential outcome $Y^a$ for four different scenarios, along with the marginal probability of exposure and the ATT. Bern($\pi$) denotes Bernoulli distribution with expectation $\pi$ and N$(\mu, 1)$ indicates Normal distribution with mean $\mu$ and variance 1.}
    \begin{tabular}{rrrrrr}
         \hline
          Scenario & $L$ &  $\mbox{logit}\{P(A=1 | L =l)\}$ & \textbf{$E(Y^a | L=l)$} & $p_1$  & $ATT$ \\
         \hline
         (i) & Bern(0.5) & $-1-2\ l$ & $- 1\ a - 1.5\ l + 1.5\ a\ l$ & 0.16 & -0.78 \\
         (ii) & Bern(0.3) & $1+0.1\ l$ & $1\ a + 1.5\ l + 0.5\ a\ l$ & 0.74 & 1.15 \\
         (iii) & N(0,1) & $1+0.1\ l$ & $1\ a+0.5\ l-1.5\ a\ l$ & 0.73 & 0.96 \\
         (iv) & N(1,1) & $1-1\ l$ & $1\ a-1.5\ l-0.5\ a\ l$ & 0.50 & 0.71 \\ 
         \hline
    \end{tabular}
    \label{tab:SimVars CT}
\end{table}

The asymptotic variances of $\widehat{ATT}$ and $\widehat{ATT}^*$ are shown in Table \ref{tab: Asy Results}. The ratio of asymptotic standard deviations, i.e., $\left(\Sigma / \Sigma^* \right)^{1/2}$, is also reported in Table \ref{tab: Asy Results} for sake of comparison with the empirical results reported in Section \ref{sec: Sim Study} below. Note from Table \ref{tab: Asy Results} that $\Sigma$ may be substantially smaller or larger than $\Sigma^*$. Thus even in large samples $\hat \Sigma$ and $\hat \Sigma^*$ may be quite different. Moreover, $\hat \Sigma^*$ will tend to yield anti-conservative inferences in scenarios (i) and (ii) and conservative inferences in scenarios (iii) and (iv). This is demonstrated empirically in the next section.


\begin{table}[ht]
\centering
\caption{The asymptotic variance of the ATT estimator when weights are unknown ($\Sigma$) and known ($\Sigma^*$),  and the ratio (Unknown / Known) of the asymptotic standard deviations (SD Ratio).}
\begin{tabular}{rrrr}
  \hline
 Scenario & $\Sigma$ & $\Sigma^*$ & SD Ratio \\
 \hline
(i) & 3.90 & 2.26 & 1.31 \\ 
  (ii) & 1.36 & 4.33 & 0.56 \\ 
  (iii) & 4.37 & 3.59 & 1.10 \\ 
  (iv) & 11.28 & 24.50 & 0.68 \\ 
   \hline
\end{tabular}
\label{tab: Asy Results} 
\end{table}

\section{Simulation Studies}\label{sec: Sim Study}

For each scenario shown in Table \ref{tab:SimVars CT}, $n=1,000$ i.i.d. copies of the variables $L,\ A,$ and $Y$ were generated for each of 1,000 datasets. For each simulated data set, $\widehat{ATT}$ was calculated using weights estimated by fitting model (\ref{eqn: wt logreg CT}). Standard errors were estimated using both $(\hat \Sigma/n)^{1/2}$ and $(\hat \Sigma^*/n)^{1/2}$. The former can be obtained with the \textit{geex} package in R~\citep{Saul2020} or the \textit{CAUSALTRT} procedure in SAS~\citep{SASCausaltrt}, and the latter is widely available in various R packages (e.g., \textit{sandwich, geeglm}) or using SAS procedures (e.g., \textit{REG}, \textit{GENMOD}). The simulation study presented here and the data analysis in the following section were conducted in R version 3.6.3~\citep{Rproj} with variance estimates computed using the \textit{geex} package. For each simulated data set Wald 95\% confidence intervals (CIs) were constructed using each SE estimate. 

Results from the simulation study are presented in Table \ref{tab: Sim Results}. In all scenarios CIs based on $\hat \Sigma$ achieved nominal coverage, whereas CIs constructed using $\hat \Sigma^*$ either under- or over-covered. These results are in agreement with the asymptotic derivations in Section \ref{sec: Mthds}. The average estimated SE ($\widehat{ASE}$) for both of the SE estimators was computed over the 1000 simulated data sets for each scenario. The $\widehat{ASE}$ ratios are reported in Table \ref{tab: Sim Results}; as expected, these ratios are very similar to the asymptotic SD ratios in Table \ref{tab: Asy Results}. 


\begin{table}[ht]
\centering
\caption{Average estimated standard error ($\widehat{ASE}$) using the stacked estimating equations (SEE) and Huber-White (HW) variance estimates, 95\% confidence interval coverage, and $\widehat{ASE}$ ratio (SEE/HW) for each simulated scenario.}
\begin{tabular}{rrrrrr}
  \hline
  & \multicolumn{2}{c}{Stacked Est Eqns} & \multicolumn{2}{c}{Huber-White} &  \\
 Scenario & $\reallywidehat{ASE}$ & Coverage & $\reallywidehat{ASE}$ & Coverage & $\reallywidehat{ASE}$ Ratio\\
 \hline
(i) & 0.062 & 0.95 & 0.048 & 0.87 & 1.31 \\ 
  (ii) & 0.037 & 0.95 & 0.066 & 1.00 & 0.56 \\ 
  (iii) & 0.066 & 0.95 & 0.060 & 0.93 & 1.10 \\ 
  (iv) & 0.106 & 0.94 & 0.157 & 1.00 & 0.67 \\ 
   \hline
\end{tabular}
\label{tab: Sim Results}
\end{table}

\section{METSIM Data Analysis}\label{sec: Data An}

\subsection{Data Characteristics and Analysis}
The METSIM cohort has been described and analyzed previously \citep{Civelek2017, Reifeis806554}. Participants of this population-based study were Finnish men aged 45-73, a subset of whom had RNA expression data recorded from an adipose tissue biopsy ($n=770$)~\citep{Laakso2017}. The exposure of interest $A$ is current smoking (yes/no) and the outcomes $Y_g,\ g=1,...,18,510$, are normalized adipose expression levels for each of 18,510 genes. Each of these gene expression outcomes will be analyzed separately. The target of inference is the ATT for each gene, i.e., the average effect of current smoking on that gene's expression in smokers. The set of variables $L$ considered sufficient for satisfying the conditional exchangeability assumption were age, alcohol consumption, body mass index (BMI), exercise level, and vegetable consumption. 

The logistic regression model (\ref{eqn: wt logreg CT}) of current smoking on the set of variables $L$ was fit to estimate the weights $\hat W_i$ for each individual. It is good practice to check that the mean of the estimated weights is close to its expected value. For the ATT weights, the expected value is $2 p_1$; see Appendix B for details. The probability $p_1$ is unknown here, but can be estimated by $\hat p_1 = \sum_i A_i /n$. For the METSIM data, the mean of the estimated weights and the estimated expected value of the weights were both 0.34. 
The IPW estimator of the ATT for each gene was computed by fitting a separate linear regression model $E(Y_g|A) = \theta_{g0} + \theta_{g1} A$ via weighted least squares using the estimated weights. The same set of individuals and weights were used for each model. Standard errors for the estimated $ATT_g$ were estimated using both $\hat \Sigma$ and $\hat \Sigma^*$. 

\subsection{Results}
\begin{figure}[H]
  \centering
  \subfloat[]{\includegraphics[width=0.45\textwidth]{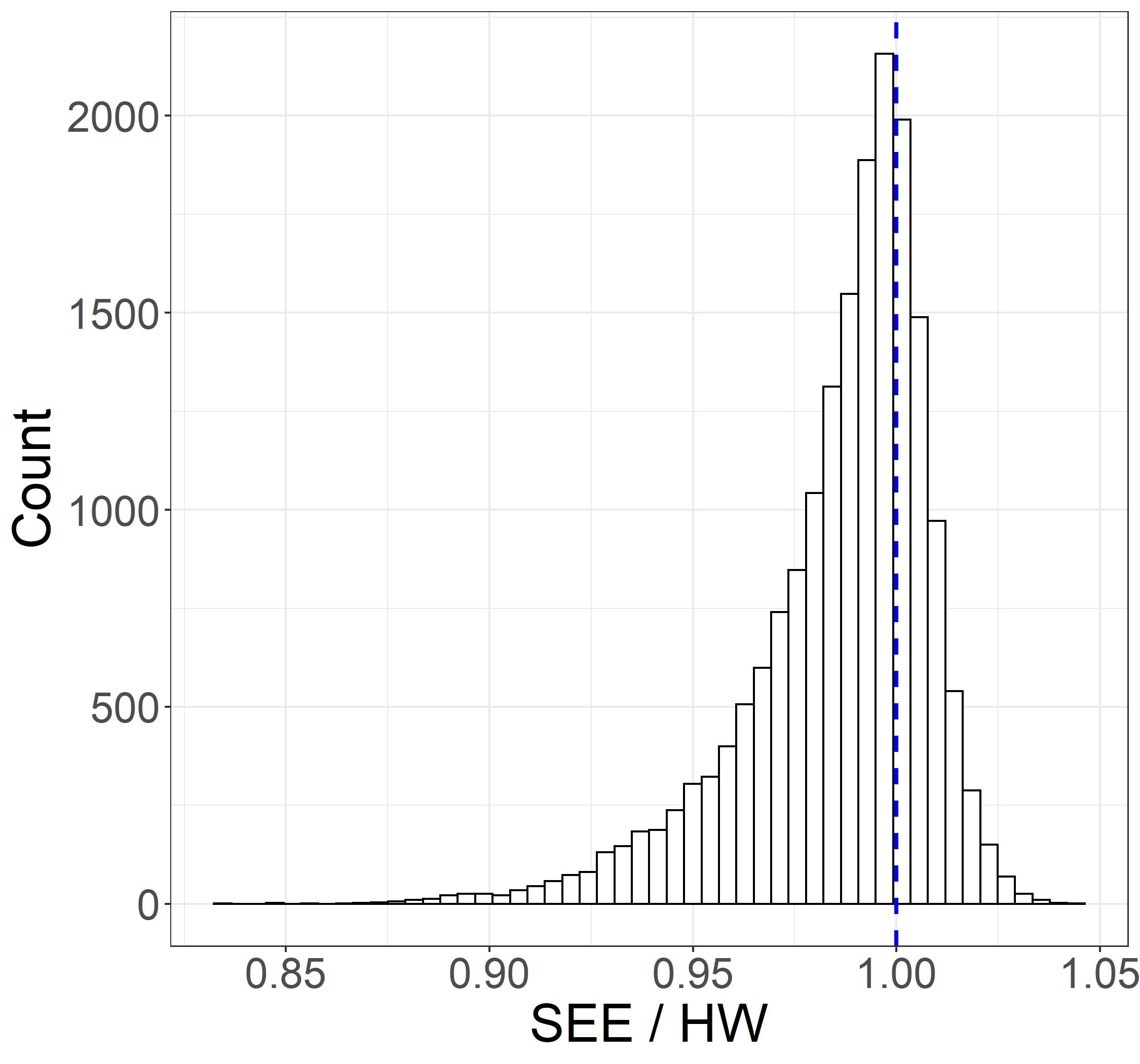} \label{subfig-1: SEratio_hist}
 } 
 \hfill
 \subfloat[]{\includegraphics[width=0.45\textwidth]{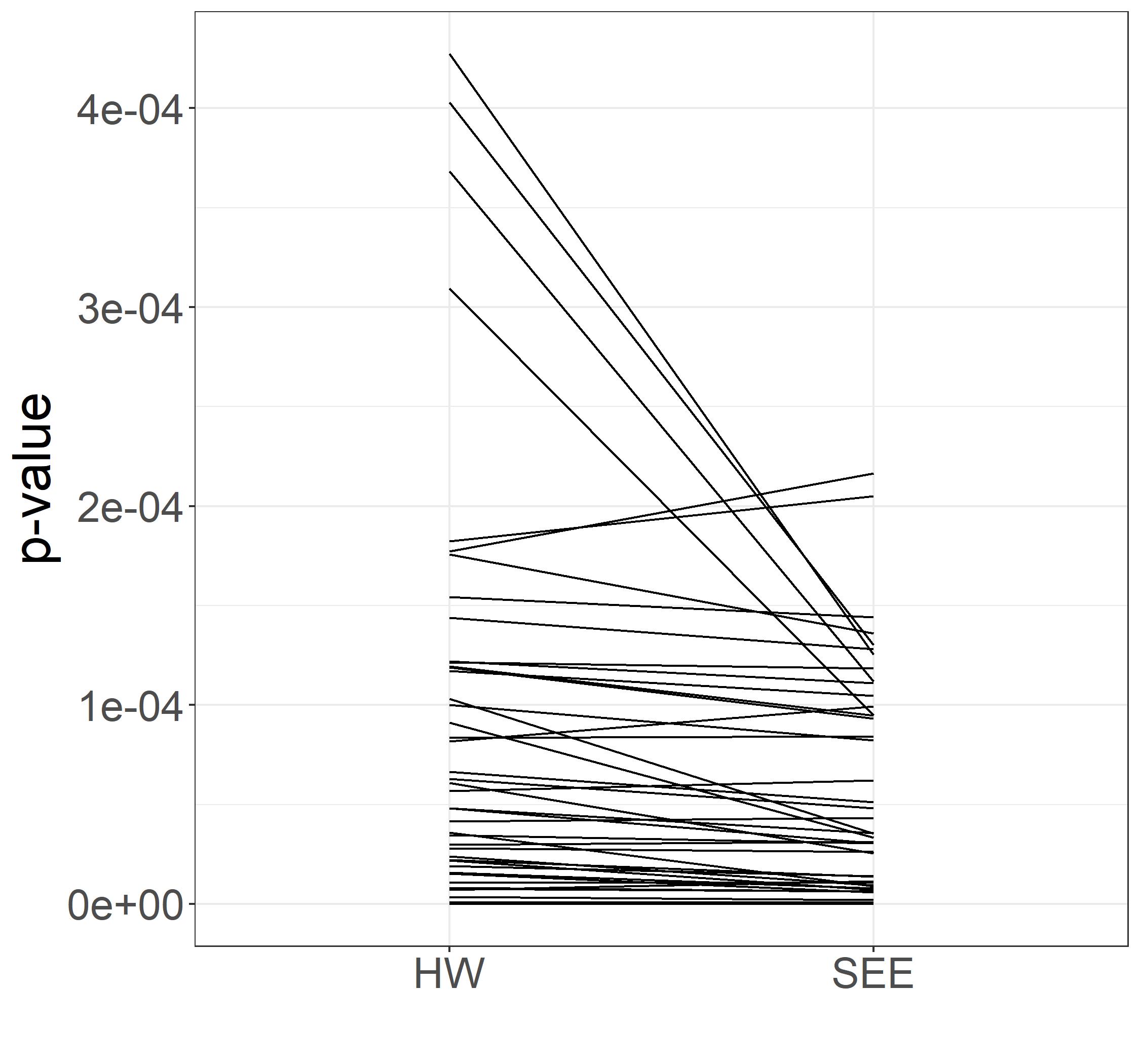} \label{subfig-2: pval_spag}
 }
 \caption{(a) Ratio of estimated SEs computed using $\hat \Sigma$ and $\hat \Sigma^*$, for the ATT of each gene in the METSIM data analysis. Vertical blue line at ratio value one denotes equality of the two SEs. (b) P-values (unadjusted) for both methods of SE estimation, for each of the top 50 genes as ranked by either method (54 genes depicted in total).}
\end{figure}

Figure \ref{subfig-1: SEratio_hist} shows the ratio of the two estimated standard errors for each of the 18,510 genes, where the vertical blue line indicates equality of the two standard error estimates. While most of the SE estimates using $\hat \Sigma^*$ were conservative (ratio less than one), there were hundreds of genes for which the estimates were anti-conservative relative to $\hat \Sigma$. The difference in SE estimates was modest for most genes, but even small differences in SE estimates can substantially affect the p-values. Figure \ref{subfig-2: pval_spag} shows raw (i.e., unadjusted for multiple testing) p-values for Wald tests of the null hypothesis $H_0: \theta_{g1} =0$ using either SE estimate. Only the 50 genes with the smallest p-values are shown. The top 50 genes as ranked by p-value differed between the two approaches, so there are 54 genes total represented in Figure \ref{subfig-2: pval_spag}. Neither $\hat \Sigma^*$ nor $\hat \Sigma$ always resulted in larger raw p-values, which aligns with the results displayed in Figure \ref{subfig-1: SEratio_hist}. Choice of standard error estimator resulted in p-values often 2-3 times larger or smaller when using $\hat \Sigma^*$ compared to $\hat \Sigma$.   

\section{Discussion}\label{sec: Discussion}

In the context of variance estimation for the ATT estimator when using IPW, assuming the weights are known can result in either a conservative or an anti-conservative variance estimate. This finding is contrary to the well known result regarding variance estimation for the ATE, namely that assuming the weights are known results in conservative variance estimates. Four simple examples are provided demonstrating that standard error estimates may be substantially larger or smaller depending on whether the weights are treated as known or estimated. Analysis of the METSIM data demonstrated how the different variance estimates can affect the ranking of genes and raw p-values.   

The variance estimator using stacked estimating equations is consistent, has a closed-form, and can be easily computed using the \textit{geex} package in R or the \textit{CAUSALTRT} procedure in SAS. R code for the asymptotic calculations in Section \ref{sec: Asy Calc} is provided in 
Appendix C, along with a workflow for analyzing a simulated data set. The bootstrap is another approach to estimating the variance of the IPW ATT estimator~\citep{ImbensGW2004, Abdia2017, Jawid2019, Were2017}, but the bootstrap estimator does not have a closed-form and can be computationally intensive. 

\section*{Acknowledgments}
The authors thank Michael Love for useful discussions of and suggestions for the manuscript. This work was supported by The Chancellor's Fellowship from The Graduate School at the University of North Carolina at Chapel Hill and NIH grant R01 AI085073. The content is solely the responsibility of the authors and does not necessarily represent the official views of the NIH.

\newpage
\bibliography{Diss_Lit_Review}
\bibliographystyle{apalike}

\newpage
\begin{appendices}
\section{IPW ATT Estimator Asymptotic Variance}\label{sec:Appx A}



Using block notation, the components of $V(\xi)$ may be expressed as 
\begin{align*}
    \mathbb{A}(\xi) &= -E \left(
    \begin{array}{cc|cc}
        \partial \psi_{\alpha_0} / \partial \alpha_0 &  \partial \psi_{\alpha_0} / \partial \alpha_1 & \partial \psi_{\alpha_0} / \partial \mu_1 & \partial \psi_{\alpha_0} / \partial \mu_0 \\
         \partial \psi_{\alpha_1} / \partial \alpha_0 & \partial \psi_{\alpha_1} / \partial \alpha_1 & \partial \psi_{\alpha_1} / \partial \mu_1 & \partial \psi_{\alpha_1} / \partial \mu_0 \\
         \hline 
         \partial \psi_1 / \partial \alpha_0 & \partial \psi_1 / \partial \alpha_1 & \partial \psi_1 / \partial \mu_1 & \partial \psi_1 / \partial \mu_0 \\
         \partial \psi_0 / \partial \alpha_0 & \partial \psi_0 / \partial \alpha_1 & \partial \psi_0 / \partial \mu_1 & \partial \psi_0 / \partial \mu_0 \\
    \end{array}
    \right) = \left( 
    \begin{array}{cc}
        a_{11} & 0_{(J+1) \times 2} \\
        a_{21} & p_1 I_2
    \end{array}
    \right) \\
    \mathbb{B}(\xi) &= E \left(
    \begin{array}{cc|cc}
        \psi_{\alpha_0}^2 &  \psi_{\alpha_0}\psi_{\alpha_1} &  \psi_{\alpha_0}\psi_{1} & \psi_{\alpha_0}\psi_{0} \\
         \psi_{\alpha_1}\psi_{\alpha_0} & \psi_{\alpha_1}^2 & \psi_{\alpha_1}\psi_{1} & \psi_{\alpha_1}\psi_{0} \\
         \hline 
         \psi_1\psi_{\alpha_0} & \psi_1\psi_{\alpha_1} & \psi_{1}^2 & \psi_{1}\psi_{0} \\
        \psi_0\psi_{\alpha_0} & \psi_0\psi_{\alpha_1} & \psi_{0}\psi_{1}  & \psi_{0}^2  
    \end{array}
    \right) = \left( \begin{array}{cc}
        b_{11} & b_{21}^T \\
        b_{21} & b_{22}
    \end{array} \right)
\end{align*}
where $\psi_{\alpha_0}, \psi_{\alpha_1}$ correspond to the score functions for the intercept and the $J$ covariates, respectively, from the logistic regression model (\ref{eqn: wt logreg CT}), and in general $0_{m \times n}$ denotes an $m \times n$ zero matrix, $0_m$ denotes a column vector of $m$ zeros, and $I_m$ denotes the $m \times m$ identity matrix. 

Next note
\begin{align*}
    \mathbb{A}(\xi)^{-1} = \left( 
    \begin{array}{cc}
        a_{11}^{-1} & 0_{(J+1) \times 2} \\
        - p_1^{-1}a_{21}a_{11}^{-1} & p_1^{-1}I_2
    \end{array}
    \right)
\end{align*}
and $a_{11}=b_{11}$ (Lemma 7.3.11, \citeauthor{casellaberger2002}, 2002), implying 
\begin{align*}
    V(\xi) &= \left( 
    \begin{array}{cc}
        a_{11}^{-1} & 0_{(J+1) \times 2} \\
         - p_1^{-1}a_{21}a_{11}^{-1} & p_1^{-1}I_2
    \end{array}
    \right)
    \left( \begin{array}{cc}
        b_{11} & b_{21}^T \\
        b_{21} & b_{22}
    \end{array} \right)
    \left( 
    \begin{array}{cc}
        a_{11}^{-1} & - p_1^{-1}a_{11}^{-1}a_{21}^T \\
        0_{(J+1)x2} &  p_1^{-1}I_2
    \end{array}
    \right) \\
    &= \left( 
    \begin{array}{cc}
        a_{11}^{-1} &  p_1^{-1}a_{11}^{-1}(-a_{21} + b_{21})^T \\
         p_1^{-1}(-a_{21} + b_{21})a_{11}^{-1} &  p_1^{-2} \{(a_{21} - b_{21})a_{11}^{-1}(a_{21} - b_{21})^T - b_{21}a_{11}^{-1}b_{21}^T + b_{22}\} \\
    \end{array}
    \right) 
\end{align*} 

By the Delta method, $\Sigma = \nabla g(\xi)^T V(\xi) \nabla g(\xi)$ where $\nabla g(\xi)^T = (0, 0_J^T, 1, -1)$. Let $c$ be the $2 \times 2$ matrix $c = (a_{21} - b_{21})a_{11}^{-1}(a_{21} - b_{21})^T - b_{21}a_{11}^{-1}b_{21}^T$, with elements $c_{11}, c_{12}, c_{21}, c_{22}$. Then
\begin{align*}
    \Sigma &= \nabla g(\xi)^T \left( 
    \begin{array}{cc}
        a_{11}^{-1} & p_1^{-1}a_{11}^{-1}(-a_{21} + b_{21})^T \\
        p_1^{-1}(-a_{21} + b_{21})a_{11}^{-1} & p_1^{-2}(c + b_{22}) \\
    \end{array}
    \right) \nabla g(\xi) \\
    &= p_1^{-2} \left[E\{A_i (Y_i^1 - \mu_1)^2\} + E\{(Y_i^0 - \mu_0)^2h(L_i; \alpha)e(L_i; \alpha)\} + c_{11} + c_{22} - c_{12} - c_{21} \right] \\
    &= \Sigma^{*} + p_1^{-2}(c_{11} + c_{22} - 2c_{12})
\end{align*}
where the last equality follows from (\ref{eqn: sig star}) and $c_{12} = c_{21}$ because $V(\xi)$ is symmetric.

Next note
\begin{align*}
    a_{21} &= \left( 
    \begin{array}{cc}
        0 & 0_J^T \\
        -E\{(1 - A_i) (Y_i - \mu_0) h(L_i; \alpha)\}  & -E\{(1 - A_i) (Y_i - \mu_0) h(L_i; \alpha) L_i^T\}
    \end{array}
    \right) \\
    b_{21} &= \left( 
    \begin{array}{cc}
        E\{A_i(Y_i - \mu_1)(1-e(L_i; \alpha))\} & E\{A_i(Y_i - \mu_1)(1-e(L_i; \alpha))L_i^T\} \\
        -E\{(1-A_i)(Y_i-\mu_0)h(L_i; \alpha)e(L_i; \alpha)\}  & -E\{(1-A_i)(Y_i-\mu_0)h(L_i; \alpha)e(L_i; \alpha)L_i^T\}
    \end{array}
    \right) 
\end{align*}
where $e(L_i; \alpha)$ and $h(L_i; \alpha)$ are as defined in the main text. It follows that
\begin{align*}
    (a_{21} - b_{21})a_{11}^{-1} 
      &= \left( \begin{array}{c}
           - E[A_i(Y_i - \mu_1)\{1-e(L_i; \alpha)\}(1, L_i^T)] \\
            -E\{(1-A_i)(Y_i - \mu_0)e(L_i; \alpha)(1, L_i^T)\}
      \end{array} \right) a_{11}^{-1}
\end{align*}

Let $m_1 = A_i(Y_i-\mu_1)\{1-e(L_i; \alpha)\}$, $m_0 = (1-A_i)(Y_i-\mu_0)e(L_i; \alpha)$, and $X_i = (1,L_i^T)$. Then
\begin{align*}
       (a_{21} - b_{21})a_{11}^{-1}(a_{21} - b_{21})^T &= \left( \begin{array}{c}
           - E(m_1 X_i) \\
            -E(m_0 X_i) 
      \end{array} \right) a_{11}^{-1} \left( \begin{array}{cc}
          - E(m_1 X_i^T) & -E(m_0 X_i^T) \\
      \end{array} \right) \\
        &= \left( \begin{array}{cc}
           E(m_1 X_i) a_{11}^{-1} E(m_1 X_i^T) & E(m_1 X_i) a_{11}^{-1} E(m_0 X_i^T) \\
           E(m_0 X_i) a_{11}^{-1} E(m_1 X_i^T) & E(m_0 X_i) a_{11}^{-1} E(m_0 X_i^T) 
      \end{array} \right)
\end{align*}
and
\begin{align*}
      b_{21}a_{11}^{-1}b_{21}^T &= \left( 
    \begin{array}{c}
        E(m_1 X_i)  \\
        -E\{m_0 h(L_i; \alpha) X_i\}
    \end{array}
    \right) a_{11}^{-1} \left( 
    \begin{array}{cc}
        E(m_1 X_i^T) & -E\{m_0 h(L_i; \alpha) X_i^T\}
    \end{array}
    \right)  \\
      &=  \left( \begin{array}{cc}
           E(m_1 X_i) a_{11}^{-1} E(m_1 X_i^T) & - E(m_1 X_i) a_{11}^{-1} E\{m_0 h(L_i; \alpha) X_i^T\} \\
           - E\{m_0 h(L_i; \alpha) X_i\} a_{11}^{-1} E(m_1 X_i^T) & E\{m_0 h(L_i; \alpha) X_i\} a_{11}^{-1} E\{m_0 h(L_i; \alpha) X_i^T\} 
      \end{array} \right)
\end{align*}
implying
\begin{align*}
c &= \left( \begin{array}{cc} 
           0  &  E(m_1 X_i) a_{11}^{-1} E[m_0 \{1+h(L_i; \alpha)\} X_i^T] \\
          E[m_0 \{1+h(L_i; \alpha)\} X_i] a_{11}^{-1} E(m_1 X_i^T)  &  E(m_0 X_i) a_{11}^{-1} E(m_0 X_i^T) - \qquad \qquad \qquad \\
          & \qquad E\{m_0 h(L_i; \alpha) X_i\} a_{11}^{-1} E\{m_0 h(L_i; \alpha) X_i^T\}
      \end{array} \right)  
\end{align*}

Assume the propensity score model (\ref{eqn: wt logreg CT}) and conditional exchangeability, the expectations above can be expressed
\begin{align*}
    E(m_1X_i) &= E[A_i(Y_i - \mu_1)\{1 - e(L_i; \alpha)\} (1,L_i^T)] \\
    &= E[A_i(Y_i^1 - \mu_1)\{1 - P(A_i=1 | L_i)\}  (1,L_i^T)] \\
    &= E_L\left[ E_{Y^1|L}(Y_i^1-\mu_1)\ P(A_i=1|L_i) \{1 - P(A_i=1|L_i)\} \  (1,L_i^T) \right]
\end{align*}
and similarly $E(m_0 X_i) = E_L\left[E_{Y^0|L}(Y_i^0 - \mu_0)\ P(A_i=1|L_i) \{1 - P(A_i=1|L_i)\} (1, L_i^T) \right]$ and \newline
\noindent $E\{m_0 h(L_i; \alpha) X_i\} = E_L\{ E_{Y^0|L}(Y_i^0 - \mu_0)\ P(A_i=1|L_i)^2 (1, L_i^T) \}$, which implies \newline 
\noindent $E\left[m_0 \{1 + h(L_i; \alpha)\} X_i\right] = E_L\{ E_{Y^0|L}(Y_i^0 - \mu_0)\ P(A_i=1|L_i) (1, L_i^T) \}$. 


Likewise, $a_{11}^{-1}=b_{11}^{-1}$ can be written
\begin{align*}
    a_{11}^{-1} &= \frac{1}{det(a_{11})} \left( \begin{array}{cc}
        E(\psi_{\alpha_1}\psi_{\alpha_1}^T) & -E(\psi_{\alpha_0} \psi_{\alpha_1})^T \\
        -E(\psi_{\alpha_0} \psi_{\alpha_1}) & E(\psi_{\alpha_0}^2)
    \end{array} \right)  
\end{align*}
where $det(a_{11}) = E(\psi_{\alpha_0}^2) E(\psi_{\alpha_1}\psi_{\alpha_1}^T) - E(\psi_{\alpha_0} \psi_{\alpha_1})E(\psi_{\alpha_0} \psi_{\alpha_1})^T $. The elements of this matrix can be expressed
\begin{align*}
    E(\psi_{\alpha_0}^2) &= E \{ A_i - 2A_iP(A_i=1|L_i) + P(A_i=1|L_i)^2 \} \\
    &= E_L [P(A_i=1|L_i)\{1 - P(A_i=1|L_i)\}]
\end{align*}
with similar derivations for $E(\psi_{\alpha_0}\psi_{\alpha_1}) = E_L [P(A_i=1|L_i)\{1 - P(A_i=1|L_i)\} L_i ]$ and $E(\psi_{\alpha_1}\psi_{\alpha_1}^T) = E_L [P(A_i=1|L_i)\{1 - P(A_i=1|L_i)\} L_iL_i^T ]$. 


Using the results above, explicit values for each element of the $c$ matrix can be calculated for given distributions of $L$, $A|L$, $Y^{0}|L$, and $Y^{1}|L$. This is demonstrated in Section \ref{sec: Asy Calc} for four example scenarios. The R code used for these calculations is included in Appendix C.

\section{Expected Value of ATT Weights}\label{sec: Appx B}

The expected value of the weights proposed by \citet{Sato2003} equals 
\begin{align*}
    E[W_i^{ATT}] &= E_{A, L}\left[A_i + (1-A_i) \frac{P(A_i=1 | L_i)}{P(A_i=0 | L_i)}\right] \\
            &= p_1 + E_L\left[E_{A|L} (1-A_i) \frac{P(A_i=1 | L_i)}{P(A_i=0 | L_i)}\right] \\
            &= p_1 + E_L\left[P(A_i=1 | L_i)\right] = 2 p_1 
\end{align*}

\section{Example R Code}\label{sec: Appx C}

The code below was written for the R environment in R version 3.6.3~\citep{Rproj}. 

\subsection*{C.1  Asymptotic
Calculations}\label{r-code-for-asymptotic-calculations}

First set the values of the parameters from scenario (i) in the main
text.

\begin{Shaded}
\begin{Highlighting}[]
\NormalTok{EL <-}\StringTok{ }\FloatTok{0.5}\NormalTok{ ; a0 <-}\StringTok{ }\OperatorTok{-}\DecValTok{1}\NormalTok{ ; a1 <-}\StringTok{ }\OperatorTok{-}\DecValTok{2}
\NormalTok{ba <-}\StringTok{ }\OperatorTok{-}\DecValTok{1}\NormalTok{ ; bL <-}\StringTok{ }\OperatorTok{-}\FloatTok{1.5}\NormalTok{ ; baL <-}\StringTok{ }\FloatTok{1.5}\NormalTok{ ; sdY <-}\StringTok{ }\FloatTok{0.5}
\end{Highlighting}
\end{Shaded}

From these defined values we can solve for the other needed quantities.

\begin{Shaded}
\begin{Highlighting}[]
\NormalTok{EA_L1 <-}\StringTok{ }\KeywordTok{exp}\NormalTok{(a0 }\OperatorTok{+}\StringTok{ }\NormalTok{a1) }\OperatorTok{/}\StringTok{ }\NormalTok{(}\DecValTok{1} \OperatorTok{+}\StringTok{ }\KeywordTok{exp}\NormalTok{(a0 }\OperatorTok{+}\StringTok{ }\NormalTok{a1))}
\NormalTok{EA_L0 <-}\StringTok{ }\KeywordTok{exp}\NormalTok{(a0) }\OperatorTok{/}\StringTok{ }\NormalTok{(}\DecValTok{1} \OperatorTok{+}\StringTok{ }\KeywordTok{exp}\NormalTok{(a0))}
  
\NormalTok{EY1_L1 <-}\StringTok{ }\NormalTok{ba }\OperatorTok{+}\StringTok{ }\NormalTok{bL }\OperatorTok{+}\StringTok{ }\NormalTok{baL }\CommentTok{#no intercept term}
\NormalTok{EY1_L0 <-}\StringTok{ }\NormalTok{ba}
\NormalTok{EY0_L1 <-}\StringTok{ }\NormalTok{bL}
\NormalTok{EY0_L0 <-}\StringTok{ }\DecValTok{0}
\NormalTok{VarY0_L <-}\StringTok{ }\NormalTok{sdY}\OperatorTok{^}\DecValTok{2}
\NormalTok{VarY1_L <-}\StringTok{ }\NormalTok{sdY}\OperatorTok{^}\DecValTok{2}
  
\NormalTok{EA <-}\StringTok{ }\NormalTok{EA_L0}\OperatorTok{*}\NormalTok{(}\DecValTok{1}\OperatorTok{-}\NormalTok{EL) }\OperatorTok{+}\StringTok{ }\NormalTok{EA_L1}\OperatorTok{*}\NormalTok{(EL)}
\NormalTok{EL_A1 <-}\StringTok{ }\NormalTok{(}\DecValTok{1}\OperatorTok{/}\NormalTok{EA) }\OperatorTok{*}\StringTok{ }\NormalTok{EA_L1 }\OperatorTok{*}\StringTok{ }\NormalTok{EL}
\NormalTok{mu0 <-}\StringTok{ }\NormalTok{bL }\OperatorTok{*}\StringTok{ }\NormalTok{EL_A1}
\NormalTok{mu1 <-}\StringTok{ }\NormalTok{ba }\OperatorTok{+}\StringTok{ }\NormalTok{bL}\OperatorTok{*}\NormalTok{EL_A1 }\OperatorTok{+}\StringTok{ }\NormalTok{baL}\OperatorTok{*}\NormalTok{EL_A1}
\NormalTok{ATT <-}\StringTok{ }\NormalTok{mu1}\OperatorTok{-}\NormalTok{mu0}
\end{Highlighting}
\end{Shaded}

These values can be plugged in to calculate the elements of the
\(a_{21},\ b_{21},\) and \(a_{11}^{-1}\) matrices.

\begin{Shaded}
\begin{Highlighting}[]
\NormalTok{## Calculate required expectations for (a21 - b21), }
\NormalTok{##  b21, and a11^\{-1\} matrices}
\NormalTok{a21_b21.}\DecValTok{1}\NormalTok{ <-}\StringTok{ }\NormalTok{(EY1_L0 }\OperatorTok{-}\StringTok{ }\NormalTok{mu1)}\OperatorTok{*}\NormalTok{EA_L0}\OperatorTok{*}\NormalTok{(}\DecValTok{1}\OperatorTok{-}\NormalTok{EA_L0)}\OperatorTok{*}\NormalTok{(}\DecValTok{1}\OperatorTok{-}\NormalTok{EL) }\OperatorTok{+}\StringTok{ }
\StringTok{                }\NormalTok{(EY1_L1 }\OperatorTok{-}\StringTok{ }\NormalTok{mu1)}\OperatorTok{*}\NormalTok{EA_L1}\OperatorTok{*}\NormalTok{(}\DecValTok{1}\OperatorTok{-}\NormalTok{EA_L1)}\OperatorTok{*}\NormalTok{EL}
\NormalTok{a21_b21.}\DecValTok{2}\NormalTok{ <-}\StringTok{ }\NormalTok{(EY0_L0 }\OperatorTok{-}\StringTok{ }\NormalTok{mu0)}\OperatorTok{*}\NormalTok{EA_L0}\OperatorTok{*}\NormalTok{(}\DecValTok{1}\OperatorTok{-}\NormalTok{EA_L0)}\OperatorTok{*}\NormalTok{(}\DecValTok{1}\OperatorTok{-}\NormalTok{EL) }\OperatorTok{+}\StringTok{ }
\StringTok{                }\NormalTok{(EY0_L1 }\OperatorTok{-}\StringTok{ }\NormalTok{mu0)}\OperatorTok{*}\NormalTok{EA_L1}\OperatorTok{*}\NormalTok{(}\DecValTok{1}\OperatorTok{-}\NormalTok{EA_L1)}\OperatorTok{*}\NormalTok{EL}
\NormalTok{a21_b21.}\DecValTok{3}\NormalTok{ <-}\StringTok{ }\NormalTok{(EY1_L0 }\OperatorTok{-}\StringTok{ }\NormalTok{mu1)}\OperatorTok{*}\NormalTok{EA_L0}\OperatorTok{*}\NormalTok{(}\DecValTok{1}\OperatorTok{-}\NormalTok{EA_L0)}\OperatorTok{*}\DecValTok{0}\OperatorTok{*}\NormalTok{(}\DecValTok{1}\OperatorTok{-}\NormalTok{EL) }\OperatorTok{+}\StringTok{ }
\StringTok{                }\NormalTok{(EY1_L1 }\OperatorTok{-}\StringTok{ }\NormalTok{mu1)}\OperatorTok{*}\NormalTok{EA_L1}\OperatorTok{*}\NormalTok{(}\DecValTok{1}\OperatorTok{-}\NormalTok{EA_L1)}\OperatorTok{*}\DecValTok{1}\OperatorTok{*}\NormalTok{EL}
\NormalTok{a21_b21.}\DecValTok{4}\NormalTok{ <-}\StringTok{ }\NormalTok{(EY0_L0 }\OperatorTok{-}\StringTok{ }\NormalTok{mu0)}\OperatorTok{*}\NormalTok{EA_L0}\OperatorTok{*}\NormalTok{(}\DecValTok{1}\OperatorTok{-}\NormalTok{EA_L0)}\OperatorTok{*}\DecValTok{0}\OperatorTok{*}\NormalTok{(}\DecValTok{1}\OperatorTok{-}\NormalTok{EL) }\OperatorTok{+}\StringTok{ }
\StringTok{                }\NormalTok{(EY0_L1 }\OperatorTok{-}\StringTok{ }\NormalTok{mu0)}\OperatorTok{*}\NormalTok{EA_L1}\OperatorTok{*}\NormalTok{(}\DecValTok{1}\OperatorTok{-}\NormalTok{EA_L1)}\OperatorTok{*}\DecValTok{1}\OperatorTok{*}\NormalTok{EL}

\NormalTok{a21_b21 <-}\StringTok{ }\KeywordTok{matrix}\NormalTok{(}\KeywordTok{c}\NormalTok{(}\OperatorTok{-}\NormalTok{a21_b21.}\DecValTok{1}\NormalTok{, }\OperatorTok{-}\NormalTok{a21_b21.}\DecValTok{2}\NormalTok{, }\OperatorTok{-}\NormalTok{a21_b21.}\DecValTok{3}\NormalTok{, }\OperatorTok{-}\NormalTok{a21_b21.}\DecValTok{4}\NormalTok{), }
                    \DataTypeTok{nrow=}\DecValTok{2}\NormalTok{, }\DataTypeTok{ncol=}\DecValTok{2}\NormalTok{)}
  
\NormalTok{b21.}\DecValTok{2}\NormalTok{ <-}\StringTok{ }\NormalTok{(EY0_L0 }\OperatorTok{-}\StringTok{ }\NormalTok{mu0)}\OperatorTok{*}\NormalTok{(EA_L0}\OperatorTok{^}\DecValTok{2}\NormalTok{)}\OperatorTok{*}\NormalTok{(}\DecValTok{1}\OperatorTok{-}\NormalTok{EL) }\OperatorTok{+}\StringTok{ }
\StringTok{              }\NormalTok{(EY0_L1 }\OperatorTok{-}\StringTok{ }\NormalTok{mu0)}\OperatorTok{*}\NormalTok{(EA_L1}\OperatorTok{^}\DecValTok{2}\NormalTok{)}\OperatorTok{*}\NormalTok{EL}
\NormalTok{b21.}\DecValTok{4}\NormalTok{ <-}\StringTok{ }\NormalTok{(EY0_L0 }\OperatorTok{-}\StringTok{ }\NormalTok{mu0)}\OperatorTok{*}\NormalTok{(EA_L0}\OperatorTok{^}\DecValTok{2}\NormalTok{)}\OperatorTok{*}\DecValTok{0}\OperatorTok{*}\NormalTok{(}\DecValTok{1}\OperatorTok{-}\NormalTok{EL) }\OperatorTok{+}\StringTok{ }
\StringTok{              }\NormalTok{(EY0_L1 }\OperatorTok{-}\StringTok{ }\NormalTok{mu0)}\OperatorTok{*}\NormalTok{(EA_L1}\OperatorTok{^}\DecValTok{2}\NormalTok{)}\OperatorTok{*}\DecValTok{1}\OperatorTok{*}\NormalTok{EL}
  
\NormalTok{b21 <-}\StringTok{ }\KeywordTok{matrix}\NormalTok{(}\KeywordTok{c}\NormalTok{(a21_b21.}\DecValTok{1}\NormalTok{, }\OperatorTok{-}\NormalTok{b21.}\DecValTok{2}\NormalTok{, a21_b21.}\DecValTok{3}\NormalTok{, }\OperatorTok{-}\NormalTok{b21.}\DecValTok{4}\NormalTok{), }
                \DataTypeTok{nrow=}\DecValTok{2}\NormalTok{, }\DataTypeTok{ncol=}\DecValTok{2}\NormalTok{)}
  
\NormalTok{a11_}\DecValTok{1}\NormalTok{ <-}\StringTok{ }\NormalTok{EA_L0}\OperatorTok{*}\NormalTok{(}\DecValTok{1}\OperatorTok{-}\NormalTok{EA_L0)}\OperatorTok{*}\NormalTok{(}\DecValTok{1}\OperatorTok{-}\NormalTok{EL) }\OperatorTok{+}\StringTok{ }\NormalTok{EA_L1}\OperatorTok{*}\NormalTok{(}\DecValTok{1}\OperatorTok{-}\NormalTok{EA_L1)}\OperatorTok{*}\NormalTok{EL}
\NormalTok{a11_}\DecValTok{2}\NormalTok{ <-}\StringTok{ }\NormalTok{EA_L0}\OperatorTok{*}\NormalTok{(}\DecValTok{1}\OperatorTok{-}\NormalTok{EA_L0)}\OperatorTok{*}\DecValTok{0}\OperatorTok{*}\NormalTok{(}\DecValTok{1}\OperatorTok{-}\NormalTok{EL) }\OperatorTok{+}\StringTok{ }\NormalTok{EA_L1}\OperatorTok{*}\NormalTok{(}\DecValTok{1}\OperatorTok{-}\NormalTok{EA_L1)}\OperatorTok{*}\DecValTok{1}\OperatorTok{*}\NormalTok{EL}
\NormalTok{a11_}\DecValTok{3}\NormalTok{ <-}\StringTok{ }\NormalTok{EA_L0}\OperatorTok{*}\NormalTok{(}\DecValTok{1}\OperatorTok{-}\NormalTok{EA_L0)}\OperatorTok{*}\NormalTok{(}\DecValTok{0}\OperatorTok{^}\DecValTok{2}\NormalTok{)}\OperatorTok{*}\NormalTok{(}\DecValTok{1}\OperatorTok{-}\NormalTok{EL) }\OperatorTok{+}\StringTok{ }\NormalTok{EA_L1}\OperatorTok{*}\NormalTok{(}\DecValTok{1}\OperatorTok{-}\NormalTok{EA_L1)}\OperatorTok{*}\NormalTok{(}\DecValTok{1}\OperatorTok{^}\DecValTok{2}\NormalTok{)}\OperatorTok{*}\NormalTok{EL}
  
\NormalTok{a11 <-}\StringTok{ }\KeywordTok{matrix}\NormalTok{(}\KeywordTok{c}\NormalTok{(a11_}\DecValTok{1}\NormalTok{, a11_}\DecValTok{2}\NormalTok{, a11_}\DecValTok{2}\NormalTok{, a11_}\DecValTok{3}\NormalTok{), }
                \DataTypeTok{nrow=}\DecValTok{2}\NormalTok{, }\DataTypeTok{ncol=}\DecValTok{2}\NormalTok{)}
\NormalTok{a11_inv <-}\StringTok{ }\KeywordTok{solve}\NormalTok{(a11)}
\end{Highlighting}
\end{Shaded}

What remains is simply using matrix algebra to calculate values of the
constant and \(\Sigma^*\).

\begin{Shaded}
\begin{Highlighting}[]
\NormalTok{## Calculate constant}
\NormalTok{c <-}\StringTok{ }\NormalTok{a21_b21 }\OperatorTok{%*%}\StringTok{ }\NormalTok{a11_inv }\OperatorTok{%*%}\StringTok{ }\KeywordTok{t}\NormalTok{(a21_b21) }\OperatorTok{-}\StringTok{ }\NormalTok{b21 }\OperatorTok{%*%}\StringTok{ }\NormalTok{a11_inv }\OperatorTok{%*%}\StringTok{ }\KeywordTok{t}\NormalTok{(b21)}
\NormalTok{c_scaled <-}\StringTok{ }\NormalTok{(}\DecValTok{1}\OperatorTok{/}\NormalTok{EA}\OperatorTok{^}\DecValTok{2}\NormalTok{)}\OperatorTok{*}\NormalTok{c  }\CommentTok{# (1/P(A=1)^2) * c}
\NormalTok{gg <-}\StringTok{ }\KeywordTok{cbind}\NormalTok{(}\KeywordTok{c}\NormalTok{(}\DecValTok{1}\NormalTok{, }\OperatorTok{-}\DecValTok{1}\NormalTok{))}
  
\NormalTok{constant <-}\StringTok{ }\KeywordTok{t}\NormalTok{(gg) }\OperatorTok{%*%}\StringTok{ }\NormalTok{c_scaled }\OperatorTok{%*%}\StringTok{ }\NormalTok{gg}
  
\NormalTok{## Calculate Sigma* and Sigma}
\NormalTok{EY0_mu02_L0 <-}\StringTok{ }\NormalTok{(VarY0_L }\OperatorTok{+}\StringTok{ }\NormalTok{EY0_L0}\OperatorTok{^}\DecValTok{2}\NormalTok{) }\OperatorTok{-}\StringTok{ }\DecValTok{2}\OperatorTok{*}\NormalTok{mu0}\OperatorTok{*}\NormalTok{EY0_L0 }\OperatorTok{+}\StringTok{ }\NormalTok{mu0}\OperatorTok{^}\DecValTok{2}
\NormalTok{EY0_mu02_L1 <-}\StringTok{ }\NormalTok{(VarY0_L }\OperatorTok{+}\StringTok{ }\NormalTok{EY0_L1}\OperatorTok{^}\DecValTok{2}\NormalTok{) }\OperatorTok{-}\StringTok{ }\DecValTok{2}\OperatorTok{*}\NormalTok{mu0}\OperatorTok{*}\NormalTok{EY0_L1 }\OperatorTok{+}\StringTok{ }\NormalTok{mu0}\OperatorTok{^}\DecValTok{2}
\NormalTok{EY1_mu12_L0 <-}\StringTok{ }\NormalTok{(VarY1_L }\OperatorTok{+}\StringTok{ }\NormalTok{EY1_L0}\OperatorTok{^}\DecValTok{2}\NormalTok{) }\OperatorTok{-}\StringTok{ }\DecValTok{2}\OperatorTok{*}\NormalTok{mu1}\OperatorTok{*}\NormalTok{EY1_L0 }\OperatorTok{+}\StringTok{ }\NormalTok{mu1}\OperatorTok{^}\DecValTok{2}
\NormalTok{EY1_mu12_L1 <-}\StringTok{ }\NormalTok{(VarY1_L }\OperatorTok{+}\StringTok{ }\NormalTok{EY1_L1}\OperatorTok{^}\DecValTok{2}\NormalTok{) }\OperatorTok{-}\StringTok{ }\DecValTok{2}\OperatorTok{*}\NormalTok{mu1}\OperatorTok{*}\NormalTok{EY1_L1 }\OperatorTok{+}\StringTok{ }\NormalTok{mu1}\OperatorTok{^}\DecValTok{2}
    
\NormalTok{b22_}\DecValTok{1}\NormalTok{ <-}\StringTok{ }\NormalTok{(EA_L0}\OperatorTok{*}\NormalTok{EY1_mu12_L0)}\OperatorTok{*}\NormalTok{(}\DecValTok{1}\OperatorTok{-}\NormalTok{EL) }\OperatorTok{+}\StringTok{ }\NormalTok{(EA_L1}\OperatorTok{*}\NormalTok{EY1_mu12_L1)}\OperatorTok{*}\NormalTok{EL   }
\NormalTok{b22_}\DecValTok{2}\NormalTok{ <-}\StringTok{ }\NormalTok{((EA_L0}\OperatorTok{^}\DecValTok{2}\OperatorTok{/}\NormalTok{(}\DecValTok{1}\OperatorTok{-}\NormalTok{EA_L0))}\OperatorTok{*}\NormalTok{EY0_mu02_L0)}\OperatorTok{*}\NormalTok{(}\DecValTok{1}\OperatorTok{-}\NormalTok{EL) }\OperatorTok{+}\StringTok{ }
\StringTok{            }\NormalTok{((EA_L1}\OperatorTok{^}\DecValTok{2}\OperatorTok{/}\NormalTok{(}\DecValTok{1}\OperatorTok{-}\NormalTok{EA_L1))}\OperatorTok{*}\NormalTok{EY0_mu02_L1)}\OperatorTok{*}\NormalTok{EL}
  
\NormalTok{Sig_star <-}\StringTok{ }\NormalTok{(b22_}\DecValTok{1} \OperatorTok{+}\StringTok{ }\NormalTok{b22_}\DecValTok{2}\NormalTok{)}\OperatorTok{/}\NormalTok{(EA}\OperatorTok{^}\DecValTok{2}\NormalTok{)}
\NormalTok{Sig <-}\StringTok{ }\NormalTok{Sig_star }\OperatorTok{+}\StringTok{ }\NormalTok{constant}

\NormalTok{df <-}\StringTok{ }\KeywordTok{data.frame}\NormalTok{(}\KeywordTok{cbind}\NormalTok{(ATT, constant, Sig_star, Sig))}
\KeywordTok{colnames}\NormalTok{(df) <-}\StringTok{ }\KeywordTok{c}\NormalTok{(}\StringTok{"ATT"}\NormalTok{, }\StringTok{"Constant"}\NormalTok{, }\StringTok{"Sigma^*"}\NormalTok{, }\StringTok{"Sigma"}\NormalTok{)}
\KeywordTok{print}\NormalTok{(df)}
\end{Highlighting}
\end{Shaded}

\begin{verbatim}
##          ATT Constant  Sigma^*    Sigma
## 1 -0.7751385 1.635956 2.263171 3.899128
\end{verbatim}

These results align with those presented in Table 2.

\subsection*{C.2  Simulated Data
Analysis}\label{r-code-for-simulated-data-analysis}

Using the population parameters defined above, we can simulate an
example data set of 1000 individuals. After generating \(L,\ A,\) and
\(Y\), the ATT weights are computed as in the main text.

\begin{Shaded}
\begin{Highlighting}[]
\KeywordTok{set.seed}\NormalTok{(}\DecValTok{42}\NormalTok{)}
\NormalTok{n <-}\StringTok{ }\DecValTok{1000}
  
\NormalTok{L <-}\StringTok{ }\KeywordTok{rbinom}\NormalTok{(n, }\DecValTok{1}\NormalTok{, }\DataTypeTok{prob =}\NormalTok{ EL)}
\NormalTok{lp <-}\StringTok{ }\KeywordTok{exp}\NormalTok{(a0 }\OperatorTok{+}\StringTok{ }\NormalTok{a1}\OperatorTok{*}\NormalTok{L)}
\NormalTok{A <-}\StringTok{ }\KeywordTok{rbinom}\NormalTok{(n, }\DataTypeTok{size =} \DecValTok{1}\NormalTok{, }\DataTypeTok{prob =}\NormalTok{ lp}\OperatorTok{/}\NormalTok{(}\DecValTok{1}\OperatorTok{+}\NormalTok{lp))}
\NormalTok{Y <-}\StringTok{ }\KeywordTok{rnorm}\NormalTok{(n, }\DataTypeTok{mean =}\NormalTok{ ba}\OperatorTok{*}\NormalTok{A }\OperatorTok{+}\StringTok{ }\NormalTok{bL}\OperatorTok{*}\NormalTok{L }\OperatorTok{+}\StringTok{ }\NormalTok{baL}\OperatorTok{*}\NormalTok{A}\OperatorTok{*}\NormalTok{L, }\DataTypeTok{sd =}\NormalTok{ sdY)}
  
\NormalTok{psmod <-}\StringTok{ }\KeywordTok{glm}\NormalTok{(A }\OperatorTok{~}\StringTok{ }\NormalTok{L, }\DataTypeTok{family =} \KeywordTok{binomial}\NormalTok{(}\DataTypeTok{link =} \StringTok{"logit"}\NormalTok{))}
\NormalTok{wt.att <-}\StringTok{ }\KeywordTok{ifelse}\NormalTok{(A }\OperatorTok{==}\StringTok{ }\DecValTok{0}\NormalTok{, }\KeywordTok{exp}\NormalTok{(psmod}\OperatorTok{$}\NormalTok{linear.predictors), }\DecValTok{1}\NormalTok{)}
  
\NormalTok{dat <-}\StringTok{ }\KeywordTok{data.frame}\NormalTok{(}\KeywordTok{cbind}\NormalTok{(L, A, Y, wt.att))}
\end{Highlighting}
\end{Shaded}

The following are helper functions defined for use within the
\texttt{geex} function \texttt{m\_estimate}, which will allow us to
compute the standard errors for the stacked estimating equations (SEE)
and robust (Huber-White) variance estimators.

\begin{Shaded}
\begin{Highlighting}[]
\NormalTok{estfun <-}\StringTok{ }\ControlFlowTok{function}\NormalTok{(data, model)\{}
\NormalTok{  L <-}\StringTok{ }\KeywordTok{model.matrix}\NormalTok{(model, }\DataTypeTok{data=}\NormalTok{data)}
\NormalTok{  A <-}\StringTok{ }\KeywordTok{model.response}\NormalTok{(}\KeywordTok{model.frame}\NormalTok{(model, }\DataTypeTok{data=}\NormalTok{data))}
\NormalTok{  Y <-}\StringTok{ }\NormalTok{data}\OperatorTok{$}\NormalTok{Y}
  
  \ControlFlowTok{function}\NormalTok{(theta)\{}
\NormalTok{    p  <-}\StringTok{ }\KeywordTok{length}\NormalTok{(theta)}
\NormalTok{    p1 <-}\StringTok{ }\KeywordTok{length}\NormalTok{(}\KeywordTok{coef}\NormalTok{(model))}
\NormalTok{    lp  <-}\StringTok{ }\NormalTok{L }\OperatorTok{%*%}\StringTok{ }\NormalTok{theta[}\DecValTok{1}\OperatorTok{:}\NormalTok{p1]}
\NormalTok{    rho <-}\StringTok{ }\KeywordTok{plogis}\NormalTok{(lp)}
    
\NormalTok{    IPW <-}\StringTok{ }\KeywordTok{ifelse}\NormalTok{(A }\OperatorTok{==}\StringTok{ }\DecValTok{1}\NormalTok{, }\DecValTok{1}\NormalTok{, }\KeywordTok{exp}\NormalTok{(lp))}
    
\NormalTok{    score_eqns <-}\StringTok{ }\KeywordTok{apply}\NormalTok{(L, }\DecValTok{2}\NormalTok{, }\ControlFlowTok{function}\NormalTok{(x) }\KeywordTok{sum}\NormalTok{((A }\OperatorTok{-}\StringTok{ }\NormalTok{rho) }\OperatorTok{*}\StringTok{ }\NormalTok{x))}
\NormalTok{    ce1 <-}\StringTok{ }\NormalTok{IPW}\OperatorTok{*}\NormalTok{(A}\OperatorTok{==}\DecValTok{1}\NormalTok{)}\OperatorTok{*}\NormalTok{(Y }\OperatorTok{-}\StringTok{ }\NormalTok{theta[p}\OperatorTok{-}\DecValTok{1}\NormalTok{]) }
\NormalTok{    ce0 <-}\StringTok{ }\NormalTok{IPW}\OperatorTok{*}\NormalTok{(A}\OperatorTok{==}\DecValTok{0}\NormalTok{)}\OperatorTok{*}\NormalTok{(Y }\OperatorTok{-}\StringTok{ }\NormalTok{theta[p]) }
    
    \KeywordTok{c}\NormalTok{(score_eqns,}
\NormalTok{      ce1,}
\NormalTok{      ce0)}
\NormalTok{  \}}
\NormalTok{\}}

\NormalTok{estfun_nolr <-}\StringTok{ }\ControlFlowTok{function}\NormalTok{(data)\{}
\NormalTok{  A <-}\StringTok{ }\NormalTok{data}\OperatorTok{$}\NormalTok{A}
\NormalTok{  Y <-}\StringTok{ }\NormalTok{data}\OperatorTok{$}\NormalTok{Y}
\NormalTok{  IPW <-}\StringTok{ }\NormalTok{data}\OperatorTok{$}\NormalTok{wt.att}

  \ControlFlowTok{function}\NormalTok{(theta)\{}
\NormalTok{    ce1 <-}\StringTok{ }\NormalTok{IPW}\OperatorTok{*}\NormalTok{(A}\OperatorTok{==}\DecValTok{1}\NormalTok{)}\OperatorTok{*}\NormalTok{(Y }\OperatorTok{-}\StringTok{ }\NormalTok{theta[}\DecValTok{1}\NormalTok{])}
\NormalTok{    ce0 <-}\StringTok{ }\NormalTok{IPW}\OperatorTok{*}\NormalTok{(A}\OperatorTok{==}\DecValTok{0}\NormalTok{)}\OperatorTok{*}\NormalTok{(Y }\OperatorTok{-}\StringTok{ }\NormalTok{theta[}\DecValTok{2}\NormalTok{])  }
    
    \KeywordTok{c}\NormalTok{(ce1,}
\NormalTok{      ce0)}
\NormalTok{  \}}
\NormalTok{\}}
\end{Highlighting}
\end{Shaded}

Fitting the weighted linear regression model yields the estimated
counterfactual means, from which we can compute the estimated ATT.

\begin{Shaded}
\begin{Highlighting}[]
\NormalTok{fit <-}\StringTok{ }\KeywordTok{geeglm}\NormalTok{(Y }\OperatorTok{~}\StringTok{ }\NormalTok{A, }\DataTypeTok{data =}\NormalTok{ dat, }\DataTypeTok{std.err =} \StringTok{'san.se'}\NormalTok{, }
                \DataTypeTok{weights =}\NormalTok{ wt.att, }\DataTypeTok{id=}\DecValTok{1}\OperatorTok{:}\KeywordTok{nrow}\NormalTok{(dat), }
                \DataTypeTok{corstr=}\StringTok{"independence"}\NormalTok{)}
\NormalTok{mu1_hat <-}\StringTok{ }\KeywordTok{mean}\NormalTok{(fit}\OperatorTok{$}\NormalTok{fitted.values[fit}\OperatorTok{$}\NormalTok{dat}\OperatorTok{$}\NormalTok{A}\OperatorTok{==}\DecValTok{1}\NormalTok{])}
\NormalTok{mu0_hat <-}\StringTok{ }\KeywordTok{mean}\NormalTok{(fit}\OperatorTok{$}\NormalTok{fitted.values[fit}\OperatorTok{$}\NormalTok{dat}\OperatorTok{$}\NormalTok{A}\OperatorTok{==}\DecValTok{0}\NormalTok{])}
  
\NormalTok{ATT_Est <-}\StringTok{ }\NormalTok{fit}\OperatorTok{$}\NormalTok{coefficients[}\DecValTok{2}\NormalTok{]  }\CommentTok{# = mu1_hat - mu0_hat}
\end{Highlighting}
\end{Shaded}

Finally, the \texttt{geex} package is used to estimate the SEs of the
estimated ATT using both the SEE and the Huber-White estimators. The
Huber-White SEs are also computed with the \texttt{geeglm} function to
check the output from \texttt{m\_estimate}.

\begin{Shaded}
\begin{Highlighting}[]
\NormalTok{## Accounting for weight estimation}
\NormalTok{results <-}\StringTok{ }\KeywordTok{m_estimate}\NormalTok{( }
              \DataTypeTok{estFUN =}\NormalTok{ estfun, }
              \DataTypeTok{data  =}\NormalTok{ dat, }
              \DataTypeTok{roots =} \KeywordTok{c}\NormalTok{(}\KeywordTok{coef}\NormalTok{(psmod), mu1_hat, mu0_hat),}
              \DataTypeTok{compute_roots =} \OtherTok{FALSE}\NormalTok{,}
              \DataTypeTok{outer_args =} \KeywordTok{list}\NormalTok{(}\DataTypeTok{model =}\NormalTok{ psmod))}
  
\NormalTok{## b22 + [1/P(A=1)^2]c}
\NormalTok{vcov_sEE <-}\StringTok{ }\KeywordTok{vcov}\NormalTok{(results)[}\DecValTok{3}\OperatorTok{:}\DecValTok{4}\NormalTok{, }\DecValTok{3}\OperatorTok{:}\DecValTok{4}\NormalTok{]}
  
\NormalTok{## Assuming weights are known}
\NormalTok{results_nolr <-}\StringTok{ }\KeywordTok{m_estimate}\NormalTok{( }
                    \DataTypeTok{estFUN =}\NormalTok{ estfun_nolr, }
                    \DataTypeTok{data  =}\NormalTok{ dat, }
                    \DataTypeTok{roots =} \KeywordTok{c}\NormalTok{(mu1_hat, mu0_hat),}
                    \DataTypeTok{compute_roots =} \OtherTok{FALSE}\NormalTok{)}
  
\NormalTok{## b22}
\NormalTok{vcov_GEE <-}\StringTok{ }\KeywordTok{vcov}\NormalTok{(results_nolr)}
  
\NormalTok{## Robust (Huber-White) Variance from geeglm for comparison}
\NormalTok{vcov_geeglm <-}\StringTok{ }\NormalTok{(}\KeywordTok{summary}\NormalTok{(fit)}\OperatorTok{$}\NormalTok{coefficients[}\DecValTok{2}\NormalTok{,}\DecValTok{2}\NormalTok{])}\OperatorTok{^}\DecValTok{2}
  

\NormalTok{Sig_est <-}\StringTok{ }\KeywordTok{t}\NormalTok{(gg) }\OperatorTok{%*%}\StringTok{ }\NormalTok{vcov_sEE }\OperatorTok{%*%}\StringTok{ }\NormalTok{gg}
\NormalTok{Sig_star_est <-}\StringTok{ }\KeywordTok{t}\NormalTok{(gg) }\OperatorTok{%*%}\StringTok{ }\NormalTok{vcov_GEE }\OperatorTok{%*%}\StringTok{ }\NormalTok{gg}

\NormalTok{df <-}\StringTok{ }\KeywordTok{data.frame}\NormalTok{(}\KeywordTok{cbind}\NormalTok{(ATT_Est, }\KeywordTok{sqrt}\NormalTok{(Sig_est), }
                       \KeywordTok{sqrt}\NormalTok{(Sig_star_est), }\KeywordTok{sqrt}\NormalTok{(vcov_geeglm)))}
\KeywordTok{colnames}\NormalTok{(df) <-}\StringTok{ }\KeywordTok{c}\NormalTok{(}\StringTok{"Est ATT"}\NormalTok{, }\StringTok{"Est SEE SE"}\NormalTok{, }
                  \StringTok{"Est HW SE (geex)"}\NormalTok{, }\StringTok{"Est HW SE (geeglm)"}\NormalTok{)}
\KeywordTok{print}\NormalTok{(df)}
\end{Highlighting}
\end{Shaded}

\begin{verbatim}
##      Est ATT Est SEE SE Est HW SE (geex) Est HW SE (geeglm)
## A -0.7543794 0.05830972       0.04407246         0.04407246
\end{verbatim}

As you can see, the SE estimates from \texttt{geeglm} and from
\texttt{geex} when weights are assumed known are the same. All estimates
resemble the results presented in Table 3, but do not match exactly
since this code was only run on one example data set and the Table 3
results are averaged over 1000 data sets.

Note that when performing the analysis for a large number of simulated
data sets or, e.g., a large genomics data set such as METSIM with hundreds
or thousands of individuals and outcomes, there is a practical need to
run the code for analyzing these data sets simultaneously on a compute
cluster.
\end{appendices}

\end{document}